\documentclass[english,aps,preprint]{revtex4}
\usepackage{amssymb}
\usepackage[T1]{fontenc}
\usepackage[latin1]{inputenc}
\usepackage{amsmath}
\usepackage{graphicx}
\makeatletter

\usepackage{subfigure}
\begin{document}

\title{Spin-torque oscillator based on tilted magnetization of the fixed layer}

\author{Yan Zhou}

\email{zhouyan@kth.se}

\affiliation{Department of Microelectronics and Applied Physics,
Royal Institute of Technology, Electrum 229, 164 40 Kista, Sweden}

\author{C. L. Zha}

\affiliation{Department of Microelectronics and Applied Physics,
Royal Institute of Technology, Electrum 229, 164 40 Kista, Sweden}

\author{S. Bonetti}

\affiliation{Department of Microelectronics and Applied Physics,
Royal Institute of Technology, Electrum 229, 164 40 Kista, Sweden}

\author{J. Persson}

\affiliation{Department of Microelectronics and Applied Physics,
Royal Institute of Technology, Electrum 229, 164 40 Kista, Sweden}

\author{Johan Åkerman}

\email{akerman1@kth.se}

\affiliation{Department of Microelectronics and Applied Physics,
Royal Institute of Technology, Electrum 229, 164 40 Kista, Sweden}

\date{\today}

\begin{abstract}
The spin torque oscillator (STO), where the magnetization of the
fixed layer is tilted out of the film plane, is capable of strong
microwave signal generation in zero magnetic field. Through
numerical simulations of the Landau-Lifshitz-Gilbert-Slonczewski
equations, within a macro-spin approximation, we study the microwave
signal generation as a function of drive current for two realistic
tilt angles. The tilt magnetization of the fixed layer can be
achieved by using a material with high out-of-plane
magnetocrystalline anisotropy, such as $\emph{L1}_{0}$ FePt.

\end{abstract}
\maketitle Broadband microwave oscillators, such as e.g. the Yttrium
Iron Garnet (YIG) oscillator, play an important role in
communication, radar applications and high-precision
instrumentation. The two major drawbacks of the YIG oscillator is
its bulk nature (e.g. 1 mm YIG crystal spheres), which foils any
attempt of monolithic integration, and its magnetic tuning, which is
both complicated and consumes high power. The Spin Torque Oscillator
\cite{Slonczewski1996,Berger1996,Katine2000,Kiselev2003,Rippard2004,Krivorotov2005aScience}
may be thought of as a modern nanoscopic analog of the YIG
oscillator: It is extremely broad band (multi octave), can achieve
high spectral purity, and is magnetically tunable with a similar
transfer function related to ferromagnetic resonance. While the STO
has significant advantages, such as easy on-chip integration and
\emph{current} tunability, it typically requires a large static
magnetic field for operation.

Various attempts have been made to realize STOs operating in the
absence of an applied magnetic field. Originally suggested by Redon
et.al. \cite{Redonpatent,Lee2005}, a perpendicularly polarized fixed
layer may drive an in-plane magnetization into an out-of-plane
precessional state even in the absence of an applied field.
Zero-field operation was indeed recently experimentally demonstrated
by Houssameddine \emph{et al.} \cite{Houssameddine2007} using an STO
with a perpendicularly polarized Co/Pt multilayer as fixed layer.
However, due to the axial symmetry of the fixed layer magnetization
and the precession, their STO requires an additional exchange biased
read-out layer on top of the free layer to break the symmetry and
generate any signal. The additional read out and exchange biasing
layer complicates the structure and its fabrication. The signal
quality is so far also quite limited compared to conventional STOs.
A different solution was first suggested by Xiao \emph{et al.}
\cite{Xiao2004} and later developed in detail by Barnas \emph{et
al.} \cite{Barnas2005}. Their suggested STO is based on a wavy
angular dependence of the spin torque, obtained by judicially
choosing free and fixed layer materials with different spin
diffusion lengths. Boulle \emph{et al.} recently fabricated such a
wavy STO and demonstrated current tunable microwave generation in
zero field \cite{Boulle2007}. Again, the output signal is quite
limited, partly caused by the associated asymmetric
magnetoresistance. A radically different approach was taken by
Pribiag \emph{et al.}, who introduced a magnetic vortex in a
\emph{thick} free layer and were able to excite zero-field
gyromagnetic precession of the magnetic vortex core through the spin
torque action from a conventional fixed layer \cite{Pribiag2007}.
While the signal quality of this vortex-STO is excellent, its
frequency range is quite limited, so far only demonstrated below 3
GHz.

\begin{figure}[ht]
\includegraphics[scale=0.9, clip=true, viewport=0.75in 4.0in 11in 8in]{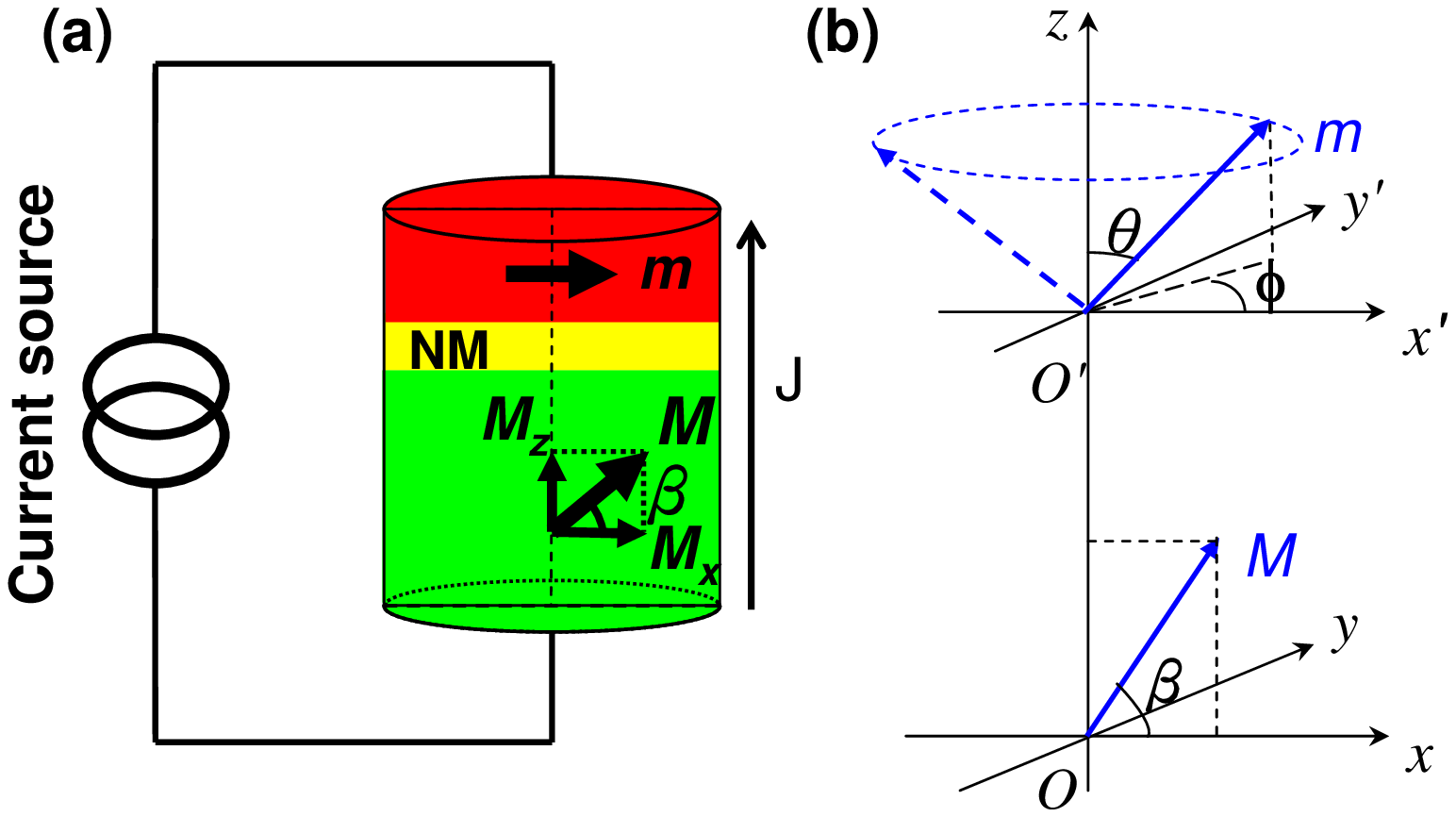}

\caption{\label{fig:TiltSTOAPL2008.Fig1Design} (a) schematic of a
TP-STO. $\textbf{M}$ is the fixed layer magnetization with tilted
orientation. The free layer magnetization $\textbf{m}$ is separated
from the fixed layer by a nonmagnetic layer (NM); (b) the coordinate
system used in this work. $\textbf{M}$ lies in the \emph{x}-\emph{z}
plane with angle $\beta$ \emph{w.r.t.} the \emph{x}-axis.}
\end{figure}

In this letter, a novel \emph{Tilted-Polarizer} STO structure
(TP-STO) has been studied where the fixed layer magnetization
($\bf{M}$) is tilted out of the film plane. In the reference frame
of the in-plane free layer magnetization ($\bf{m}$), the spin
polarization hence has both in-plane components ($p_x , p_y$) and a
component along the out-of-plane direction ($p_z$). We show that
$p_z$ can drive the free layer into precession \emph{without} the
need for an applied field, while the in-plane component $M_x$ of the
fixed layer magnetization generates a large magnetoresistance (MR),
i.e. an rf output \emph{without} the need for an additional read-out
layer. While $\bf{M}$ may have any out-of-plane direction in the
general situation, we limit our discussion to the \emph{x}-\emph{z}
plane, $\bf{M}$=($M_x, 0, M_z$)=|$\bf{M}$|($\cos \beta, 0, \sin
\beta$), and $\beta$=36$^\circ$ and 45$^\circ$ (Fig.\thickspace
\ref{fig:TiltSTOAPL2008.Fig1Design}), since these two particular
angles can be achieved using different crystallographic orientations
of FePt.

The time-evolution of the free layer magnetizaiton $\hat{m}$ is
found using the standard Landau-Lifshitz-Gilbert-Slonczewski (LLGS)
equation,
\begin{equation}
\label{Eq:LLGS}
\begin{array}{c}
\dfrac{d\hat{m}}{dt}=-\gamma\hat{m}\times{{\bf{H}}_{eff}}+\alpha\hat{m}\times\dfrac{d\hat{m}}{dt}\\
+\dfrac{\gamma}{\mu_{0}M_{S,free}}\tau,
\end{array}
\end{equation}
where $\hat{m}$ is the unit vector of the free layer magnetization,
$M_{S,free}$ its saturation magnetization, $\gamma$ the gyromagnetic
ratio, $\alpha$ the Gilbert damping parameter, and $\mu_{0}$ the
magnetic vacuum permeability. Setting the applied field to zero and
separating the effect of the demagnetizing tensor into a positive
anisotropy field along \emph{x}, and a negative out-of-plane
demagnetizing field we get $\bf{H}_{eff}$=$(H_k \hat{e}_x m_x$-$H_d
\hat{e}_z m_z)/|\bf{m}|$. We define positive current as flowing from
the fixed layer to the free layer. The quantity $\tau$ in Eq.
\ref{Eq:LLGS} is the Slonczewski spin-transfer torque density,
\begin{equation}
\label{Eq:Spintorque}
\begin{array}{c}
\tau=\eta(\varphi)\dfrac{\hbar J}{2 e d}
\hat{m}\times(\hat{m}\times\hat{M}),
\end{array}
\end{equation}
where $\varphi$ is the angle between $\hat{m}$ and $\hat{M}$,
\emph{d} is the free layer thickness, and
%
%
\begin{equation}
\label{Eq:ASSpinTorqueForm}
\begin{array}{c}
\eta(\varphi)=\dfrac{q_{+}}{A+B \cos(\varphi)}+\dfrac{q_{-}}{A-B
\cos(\varphi)}.
\end{array}
\end{equation}
where $q_{+}$, $q_{-}$, A, B are all material dependent parameters
\cite{xiaoj2005}. In our simulations below we use Cu as spacer,
Permalloy (Py) as the free layer, and FePt as the fixed layer. Due
to the lack of available parameters for the Cu/FePt interface, we
approximate $\eta(\varphi)$ in our Py/Cu/FePt stack using literature
values for Py/Cu/Co
\cite{Xiao2004,Barnas2005,Barnas2006MSE,Gmitra2007,Gmitra2006,Gmitra2006PRL,xiaoj2005}.
This approximation may be justified if a thin polarizing layer of Co
is used at the Cu/FePt interface.

We use the following generalized form for describing the angular
dependence of MR
\cite{Slonczewski2002,Krivorotov2007,Stiles2002,Urazhdin2005,Dauguet1996,Vedyayev1997},
\begin{equation}
\label{Eq:ASMRForm}
\begin{array}{c}
r=\dfrac{R(\varphi)-R_P}{R_{AP}-R_P}=
\dfrac{{1-\cos^2(\varphi/2)}}{{1+\chi\cos^2(\varphi/2)}},
\end{array}
\end{equation}
where $r$ is the reduced MR, $\chi$ is an asymmetry parameter
describing the deviation from sinusoidal angular dependence, and
$R_P$ and $R_{AP}$ denotes the resistance in the parallel and
antiparallel configurations respectively.

While both the asymmetric torque and the asymmetric
magnetoresistance are derived for in-plane spin polarizations and
magnetizations, we argue that they will still hold as long as
spin-orbit coupling is weak. While this is true for Py, it might
still be questionable for FePt due to its large magnetocrystalline
anisotropy. We argue that any deviation due to strong spin-orbit
coupling will not change the general result of our study and is
likely further weakened by our choice of a thin polarizing layer of
Co on top of FePt.

\begin{figure}[ht]
\includegraphics[scale=0.9, clip=true, viewport=3.in 0.4in 8in 7.6in]{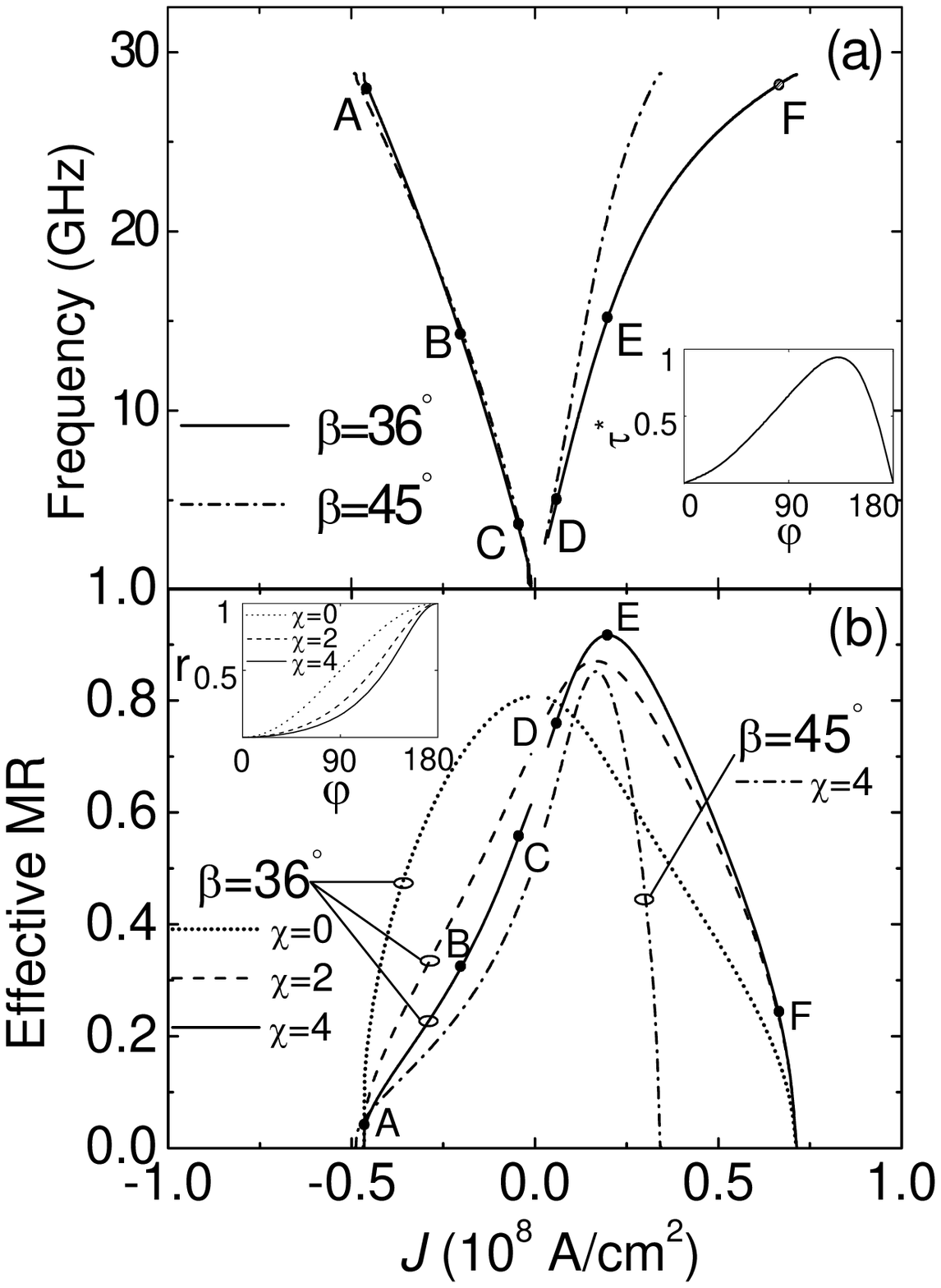}\includegraphics[scale=0.5, clip=true, viewport=3.19in -3.3in 11in 7in]{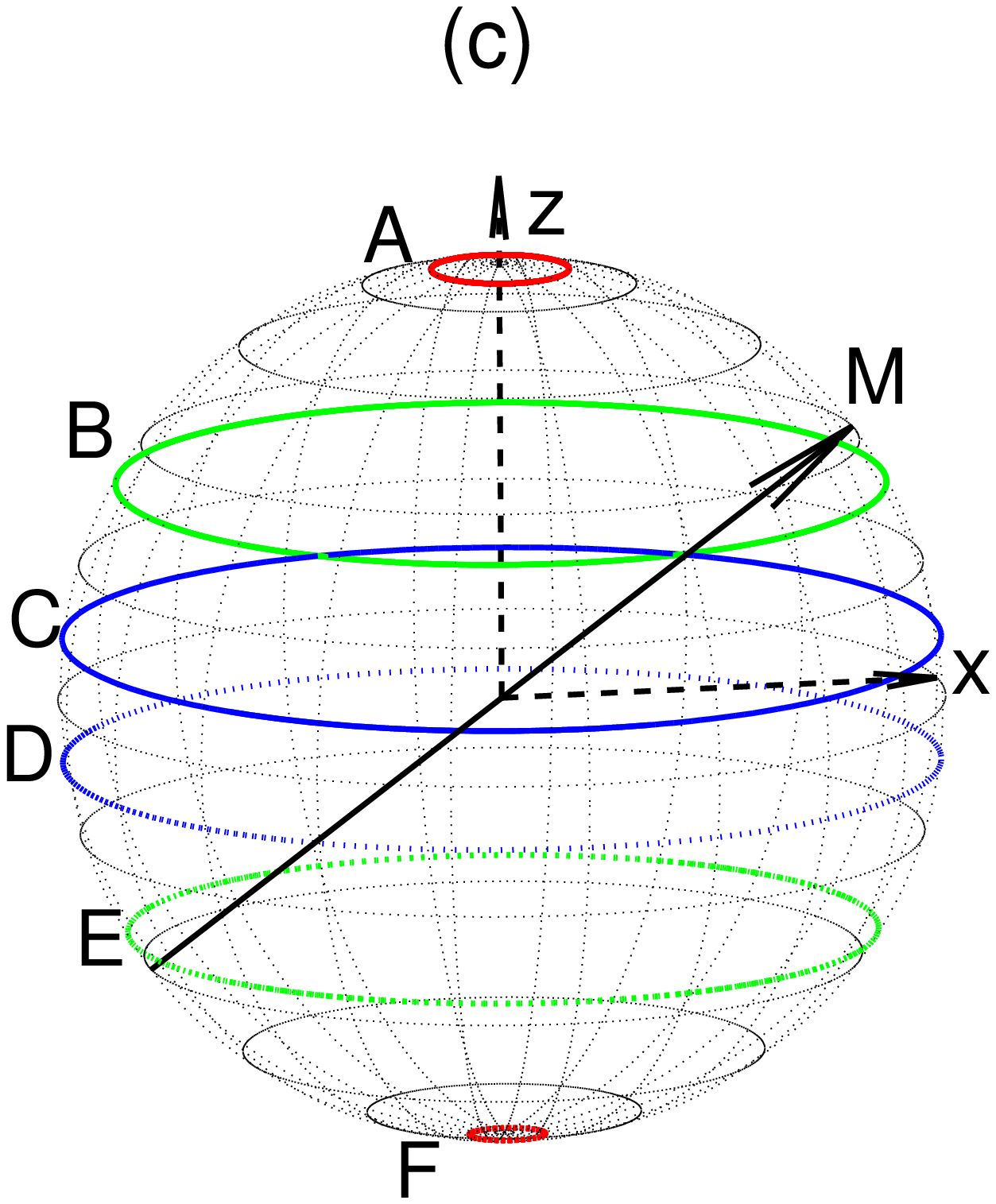}

\caption{\label{fig:TiltSTOAPL2008.Fig2} (a) Precession frequency
vs. drive current for $\beta$=36$^\circ$ (solid line) and
$\beta$=45$^\circ$ (dash dot line). Inset: Normalized spin torque
$\tau^{*}$=4$ed\tau/\hbar J$ vs. $\varphi$. (b) Effective MR vs.
\emph{J} for $\beta$=36$^\circ$(solid line) and $\beta$=45$^\circ$
(dashed line). Inset: Reduced MR vs. $\varphi$. (c) Precession
orbits on the unit sphere for different \emph{J} and
$\beta$=36$^\circ$.}
\end{figure}

Fig. \ref{fig:TiltSTOAPL2008.Fig2}a shows the precession frequency
vs. drive current density for the two selected angles. We observe
precession at both positive and negative current and the frequency
increases with the magnitude of the current density, similar to
perpendicularly polarized STOs \cite{Lee2005,Zhu2006,Jin2006}. As in
simulations for perpendicularly polarized STOs \cite{Lee2005}, the
precession starts along the equator and continues to follow
increasing latitudes of the unit sphere throughout the entire
frequency range ($f$ increases due to the increasing demagnetizing
field) until it reaches a static state at the north (south) pole for
large negative (positive) current (Fig.
\ref{fig:TiltSTOAPL2008.Fig2}c). We highlight six orbits (denoted by
A, B, C, D, E, F) which correspond to points in Fig.
\ref{fig:TiltSTOAPL2008.Fig2}a and \ref{fig:TiltSTOAPL2008.Fig2}b.
$\bf{m}$ precesses in the north hemisphere for negative \emph{J} and
in the south hemisphere for positive \emph{J} in an attempt to
allign/anti-allign with $\bf{M}$. We hence conclude that the
precession is largely dominated by the perpendicular component $p_z$
of the spin polarized current and virtually independent of the
in-plane components. The asymmetry of the dependence for different
current polarity is due to the asymmetric spin torque form as shown
in the inset of Fig. \ref{fig:TiltSTOAPL2008.Fig2}a.

Fig. \ref{fig:TiltSTOAPL2008.Fig2}b shows the effective MR
(MR$_{\mathrm{eff}}$) as a function of current density for the two
tilt angles and different choices of $\chi$. MR$_{\mathrm{eff}} J^2$
is a measure of the expected rf output where MR$_{\mathrm{eff}}$ is
the difference between the maximum and minimum resistance values
along the orbit normalized by the full $R_{AP}$-$R_{P}$. As the
precession orbit contracts with increasing $|J|$, one may expect
MR$_{\mathrm{eff}}$ to be maximum at the equator and exhibit a
monotonic decrease with increasing $|J|$. While symmetric MR
($\chi$=0) indeed yields a maximum MR$_{\mathrm{eff}}$ at the onset
of precession, the higher the MR asymmetry, the more the
MR$_{\mathrm{eff}}$ peak gets shifted to higher positive current.
For asymmetric angular dependence of MR, it is hence favorable to
precess at a finite latitude with a larger average angle
\emph{w.r.t} to $\bf{M}$. For optimal output it is consequently
desirable to use positive currents and tailor $\chi$ as to position
MR$_{\mathrm{eff}}$ in the middle of the operating frequency range.
While there are no $\chi$ values reported for NiFe/Cu/FePt, $\chi$
may range from 0 to 4 in other trilayers involving NiFe
\cite{Smith2006,Shpiro2003,Krivorotov2007,Urazhdin2005}.

\begin{figure}[ht]
\includegraphics[scale=0.6, clip=true, viewport=0.8in 0.2in 11in 11in]{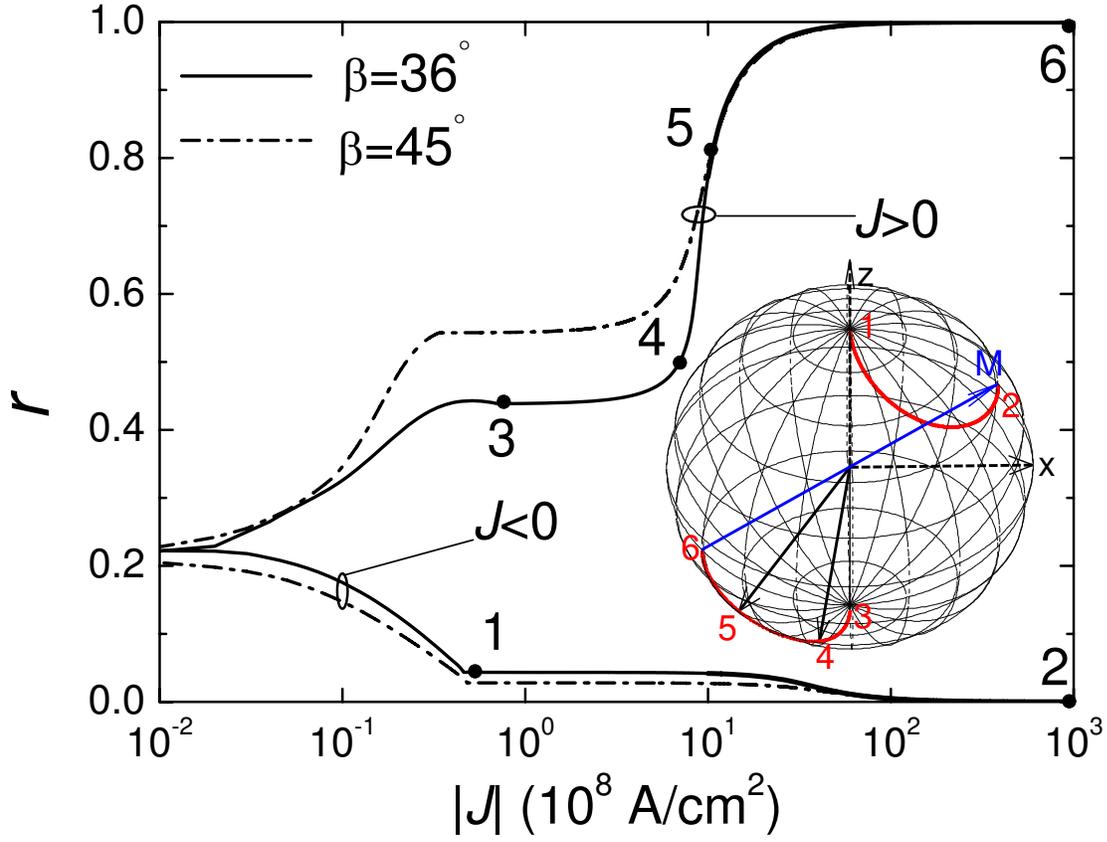}
\caption{\label{fig:TiltSTOAPL2008.Fig3.StaticOrbit}
Magnetoresistance as a function of current density. Inset: the
equilibrium states of $\hat{m}$ at different current densities when
$\beta$=36$^\circ$. 1: $J$=-0.5$\times 10^8$ A/cm$^2$; 2:
$J$=-1$\times 10^{11}$ A/cm$^2$; 3: $J$=0.75$\times 10^8$ A/cm$^2$;
4: $J$=7$\times 10^8$ A/cm$^2$; 5: $J$=1$\times 10^9$ A/cm$^2$; 6:
$J$=1$\times 10^{11}$ A/cm$^2$.}
\end{figure}

It is interesting to note that despite the large in-plane component
of the spin polarization, the initial static states are virtually
identical to the north and south poles where the spin torque and the
torque from the demagnetizing field balance each other. However, if
we further increase |\emph{J}| we expect this equilibrium point to
move towards (anti)alignment with $\hat{M}$. As shown in Fig.
\ref{fig:TiltSTOAPL2008.Fig3.StaticOrbit}, $\hat{m}$ starts out at
$\theta \approx 1^\circ$ and $177^\circ$ and then gradually follows
a curved trajectory to align with $\hat{M}$ at very large negative
current and anti-align at very large positive current. The
resistance will change accordingly and at very large currents reach
$R_{P}$ and $R_{AP}$ respectively.

There are several experimental ways to achieve easy-axis tilted hard
magnets
\cite{lubinpatent,Wang2005,Zheng2002,Klemmer2006,Zha2006,Jeong2001}.
For example, an easy axis orientation of $36^\circ$ can be achieved
by growing \emph{$L1_{0}$} (111) FePt on conventional  Si (001)
substrate \cite{Zha2006} or on MgO(111) underlayer \cite{Jeong2001}.
The $45^\circ$ orientation can be achieved by epitaxially growing an
\emph{$L1_{0}$} (101) FePt thin film on a suitable seed layer (e.g.
CrW (110) with bcc lattice) at a temperature above T=$350^\circ$C
\cite{lubinpatent}. \emph{$L1_{0}$} FePt has high magnetocrystalline
anisotropy ($K_{u}$=7$\times 10^{7}$ erg/cm$^{3}$), high saturation
magnetization ($M_{s}$=1140 emu/cm$^{3}$), and a high Curie
temperature ($T_{c}$=750 K). In both cases, a thin Co layer may be
deposited on top of the fixed layer to promote a high degree of spin
polarization. This is then followed by a standard Cu spacer and a
NiFe free layer.

In summary, the Tilted-Polarizer STO structure (TP-STO), where the
fixed layer has a tilted out-of-plane oriented magnetization, yields
the combined advantage of zero-field operation and high output
signal without the need for an additional sensing layer. Both the
precession and effective MR dependence on the driving current and
the equilibrium states of the STO can be well understood by
investigating the precession orbits of the free layer. The TP-STOs
with tilt angles $\beta$=36$^\circ$ and 45$^\circ$ can be fabricated
by using FePt with high anisotropy and tilted easy-axis.

We thank M. Stiles for useful discussions. Support from The Swedish
Foundation for strategic Research (SSF), The Swedish Research
Council (VR), and the Göran Gustafsson Foundation is gratefully
acknowledged.
\newpage

\newpage
\newpage
\newpage
\newpage
\newpage
\newpage

\end{document}